\documentclass[twocolumn,prd,nofootinbib,showpacs,preprintnumbers]{revtex4}
\usepackage{epsfig}
\usepackage{graphicx}
\usepackage{amsmath}
\usepackage{dsfont}

\newcommand{\href}[2]{#2}

\def\bD{{\bf D}}
\def\bg{{\bf g}}

\def\bL{{\bf L}}
\def\bB{{\bf B}}

\newcommand{\eg}{\emph{e.g.}}
\newcommand{\ie}{\emph{i.e.}}
\newcommand{\Ref}{Ref.}

\newcommand{\fig}{Fig.}
\newcommand{\figs}{Figs.}
\newcommand{\eq}{Eq.}

\DeclareMathOperator{\diag}{diag}

\begin{document}
\title{Nonstandard neutrino-neutrino refractive effects in dense neutrino gases}

\preprint{MPP-2008-134} 
\preprint{CERN-PH-TH/2008-223} 
\preprint{FERMILAB-PUB-08-532-A}

\author{Mattias Blennow}\email{blennow@mppmu.mpg.de}
\affiliation{Max-Planck-Institut f\"ur Physik
(Werner-Heisenberg-Institut), F\"ohringer Ring 6, 80805 M\"unchen,
Germany}

\author{Alessandro Mirizzi}\email{amirizzi@mppmu.mpg.de}
\affiliation{Max-Planck-Institut f\"ur Physik
(Werner-Heisenberg-Institut), F\"ohringer Ring 6, 80805 M\"unchen,
Germany}

\author{Pasquale D.~Serpico}\email{serpico@cern.ch}
\affiliation{Physics Department, Theory Division,
CERN, CH-1211 Geneva 23, Switzerland}
\affiliation{Center for Particle Astrophysics, Fermi National Accelerator
Laboratory, Batavia, IL 60510-0500 USA.}

\date{December 17, 2008}

\begin{abstract}
We investigate the effects of nonstandard four-fermion
neutrino-neutrino interactions on the flavor evolution of dense
neutrino gases. We find that in the regions where the
neutrino-neutrino refractive index leads to collective flavor
oscillations, the presence of new neutrino interactions can produce
flavor equilibration in both normal and inverted neutrino mass
hierarchy. In realistic supernova environments, these effects are
significant if the nonstandard neutrino-neutrino interaction strength
is comparable to the one expected in the standard case, dominating the
ordinary matter potential. However, very small nonstandard
neutrino-neutrino couplings are enough to trigger the usual collective
neutrino flavor transformations in the inverted neutrino mass
hierarchy, even if the mixing angle vanishes exactly.
\end{abstract}
\maketitle

\section{Introduction}

New neutrino interactions are predicted by several extensions of the
standard electroweak theory. These new interactions can be treated in
an effective (low-energy) framework by using four-fermion operators,
\eg, $\mathcal{O}_{\alpha\beta}\sim [\overline\nu_\alpha \gamma^\mu
  P_L\nu_\beta] [\overline f \gamma_\mu P_L f]$ with strength
$G_{\alpha\beta}$, inducing either flavor-changing ($\alpha\neq\beta$)
or diagonal but flavor non-universal neutrino transitions. Nonstandard
interactions (NSI) of neutrinos with charged fermions in matter, and
their interplay with the oscillation phenomenon, have been
investigated in many different contexts. An incomplete list includes
analyses related to the solar neutrino
problem~\cite{Guzzo:1991hi,Bergmann:2000gp,Gago:2001si,Friedland:2004pp,Guzzo:2004ue},
to the atmospheric neutrino
anomaly~\cite{Bergmann:1999pk,Fornengo:2001pm,Friedland:2004ah,Friedland:2005vy},
to supernova
neutrinos~\cite{Nunokawa:1996tg,Nunokawa:1996ve,EstebanPretel:2007yu,Mansour:1997fi,%
  Fogli:2002xj,Amanik:2004vm,Amanik:2006ad}, to primordial
neutrinos~\cite{Mangano:2006ar}, to the production or detection of
laboratory neutrinos~\cite{Grossman:1995wx,Bergmann:1998ft}, and to
future long-baseline
projects~\cite{Huber:2002bi,Bandyopadhyay:2007kx,Kopp:2008ds}. Neutrino-neutrino
interactions are even more difficult to constrain
experimentally~\cite{Masso:1994ww,Bilenky:1992xn,Bilenky:1994ma}.  In
particular, four-fermion (left-handed) neutrino-neutrino interactions
as large as the Standard Model neutral current couplings are viable
without violating astrophysical or laboratory
bounds~\cite{Bilenky:1999dn} (see also \Ref~\cite{Gavela:2008ra} for a
recent discussion on how to generate this type of interactions in a
gauge invariant way).

Dense astrophysical environments like the Early Universe,
core-collapse supernovae and accretion disks of coalescing neutron
stars are among the few systems where neutrino-neutrino interactions
play a role, via the refractive index that can affect the flavor
evolution of the system.  In particular, it has been shown that flavor
oscillations in dense neutrino gases exhibit collective phenomena
caused by neutrino-neutrino interactions~\cite{Pantaleone:1992eq,%
  Samuel:1993uw, Wong:2002fa,Abazajian:2002qx,Pastor:2008ti,%
   Qian:1995ua, Pastor:2002we, Sawyer:2005jk,%
  Fuller:2005ae, Duan:2005cp, Duan:2006an, Hannestad:2006nj,%
  Duan:2007mv, Raffelt:2007yz, EstebanPretel:2007ec, Raffelt:2007cb,%
  Raffelt:2007xt, Duan:2007fw, Fogli:2007bk, Duan:2007bt, Duan:2007sh,%
  Dasgupta:2008cd, EstebanPretel:2007yq, Dasgupta:2007ws, Duan:2008za,%
  Dasgupta:2008my,Sawyer:2008zs, Duan:2008eb, Chakraborty:2008zp,%
  Dasgupta:2008cu,EstebanPretel:2008ni,Gava:2008rp,Fogli:2008pt,Duan:2008fd,Raffelt:2008hr}.  Since physics
beyond the Standard Model can plausibly induce at least small deviations from the
standard predictions, it seems interesting to investigate the impact
of new neutrino-neutrino interactions on collective flavor
oscillations in dense neutrino gases.

We devote our work to this purpose.  In Section~\ref{EqoM}, we
generalize the equations of motion for a neutrino ensemble in the
presence of new neutrino-neutrino interactions. We present our results
for the flavor evolution of a homogeneous and isotropic gas of
neutrinos with decreasing density in Section~\ref{SinAn}, where we
find that NSI among neutrinos could completely equilibrate the flavor
content of the system in both normal and inverted neutrino mass
hierarchy. In the case of neutrinos streaming off a supernova (SN)
core, these new effects are inhibited by the strong ordinary matter
potential unless the strength of the NSI is of the same order of
magnitude as the one predicted by the Standard Model.  However, also
in the case of small values of nonstandard neutrino-neutrino
couplings, these are enough to trigger collective flavor
transformations in inverted mass hierarchy for vanishing mixing, where
in the standard case no evolution is expected.  Finally, in
Section~\ref{concl} we comment our results and conclude.

\section{Equations of motion with nonstandard neutrino-neutrino interactions}\label{EqoM}

Flavor oscillations of a homogeneous ensemble of neutrinos and
antineutrinos  are described by an equation of motion (EOM)
for each mode ${\bf p}$~\cite{Sigl:1992fn}
\begin{eqnarray}
i\dot \rho_{\bf p} &=& \left[(\Omega^0_{\bf p} + V +\Omega^S_{\bf p}), \rho_{\bf p}
\right]  \,\ , \\
-i\dot {\bar\rho}_{\bf p} &=& \left[(\Omega^0_{\bf p} - V -\Omega^S_{\bf p}), 
{\bar\rho}_{\bf p}
\right] \,\ ,
\end{eqnarray}
where $[\cdot,\cdot]$ denotes the commutator. For ultrarelativistic
neutrinos of momentum $p$, the matrix of vacuum oscillation
frequencies, expressed in the mass basis, is $\Omega^0_{\bf
  p}=\diag(m_1^2, m_2^2, m_3^2)/2p$, $m_{i=1,2,3}$ being the neutrino
masses.  At leading order, the only non-vanishing element of the
matter potential matrix in the weak interaction basis is due to the
$\nu_e$ forward scattering on background electrons, leading to $V=
\sqrt{2}G_F n_e \diag(1,0,0)$, where $G_F$ is the Fermi constant and
$n_e$ is the net electron number density.
Neutrino-neutrino self-interactions introduce an additional
contribution $\Omega^S_{\bf p}$ to the refractive energy shift. One
finds~\cite{Sigl:1992fn}
\begin{eqnarray}
\Omega^S_{\bf p} &=& \sqrt{2}G_F \int {\rm d}{\bf q}  \,\
\left(1-{\bf v}_{\bf q}\cdot{\bf v}_{\bf p}\right)
\bigg\{ G (\rho_{\bf q} -{\bar \rho}_{\bf q}) G   \nonumber \\
&  & + G\,\textrm{Tr}\bigg[
(\rho_{\bf q} -{\bar \rho}_{\bf q}) G \bigg ] \bigg\}  ,
\end{eqnarray}
where 
${\rm d}{\bf q} \equiv {\rm d}^3{\bf q} /{(2\pi)^3}$
 and ${\bf v}_{\bf p}$ is the velocity.
 The factor $(1-{\bf v}_{\bf
  q}\cdot{\bf v}_{\bf p})$ implies ``multi-angle effects'' for
neutrinos moving on different trajectories~\cite{Duan:2006an}.  In the Standard  Model,
the dimensionless coupling matrix $G$ is the identity matrix. If we consider
the possibility of new physics beyond the Standard Model, the coupling
matrix can assume a non-universal and/or non-diagonal structure.

We now focus on a two-flavor system $\{\nu_e,\nu_x\}$, where we expand all the
$2{\times}2$ matrices in the EOM in terms of the $2{\times}2$ unit matrix ${\mathds I}$ and the
Pauli matrices {\boldmath $\sigma$}. 
Explicitly, we 
define~\cite{Hannestad:2006nj}
\begin{eqnarray}\label{eq:matrices}
 \Omega_{\bf p}^0&=&{\textstyle\frac{1}{2}}
 \bigl(\omega_0\,{\mathds I}+\omega_{\bf p}\,{\bf B}
 \cdot\hbox{\boldmath$\sigma$}\bigr)\,,\nonumber\\
V&=&  {\textstyle\frac{\lambda}{2}}
 \bigl({\mathds I}+{\bf L}
 \cdot\hbox{\boldmath$\sigma$}\bigr)\,,\nonumber\\
 \rho_{\bf p}&=&{\textstyle\frac{1}{2}}
 \bigl(f_{\bf p}\,{\mathds I}+ n_{\bar \nu}{\bf P}_{\bf p}
 \cdot\hbox{\boldmath$\sigma$}\bigr)\,,\nonumber\\
\bar\rho_{\bf p}&=&{\textstyle\frac{1}{2}}
 \bigl(\bar f_{\bf p}\,{\mathds I}+{n}_{\bar \nu}\bar{\bf P}_{\bf p}
 \cdot\hbox{\boldmath$\sigma$}\bigr)\,,\nonumber\\
  G  &=&{\textstyle\frac{1}{2}}
 \bigl(g_0\,{\mathds I}+ {\bf g}
 \cdot\hbox{\boldmath$\sigma$}\bigr)\,.
\end{eqnarray}
Here, the overall neutrino (antineutrino) density is given by
$\int{\rm d}{\bf p}\,f_{\bf p}=n_\nu$ ($\int{\rm d}{\bf p}\,{\bar
  f}_{\bf p}=n_{\bar\nu}$). For simplicity and without loss of
generality, we here assume that initially only $\nu_e$ and $\bar\nu_e$
are present with an excess neutrino density of
$n_{\nu_e}=(1+\xi)\,n_{\bar\nu_e}$; in numerical examples, we shall
assume that the asymmetry parameter between neutrino species is
$\xi=0.25$~\cite{EstebanPretel:2007ec}.  The vectors ${\bf P}_{\bf p}$
and $\bar{\bf P}_{\bf p}$ are the neutrino and antineutrino
polarization vectors.  We define the total polarization vectors ${\bf
  P} = \int{\rm d}{\bf p}\,{{\bf P}_{\bf p}}$, $\bar{\bf P} = \int{\rm
  d}{\bf p}\,\bar{\bf P}_{\bf p}$, which are initially normalized such
that ${\bf P}(0)=(1+\xi){\bf e}_z$ and ${\bar{\bf P}}(0)={\bf e}_z$,
${\bf e}_z$ being the unit vector in the positive $z$-direction.
Here, we have chosen our coordinate system in such a way that a
polarization vector pointing in the positive $z$-direction represents
electron neutrinos, whereas an orientation in the negative
$z$-direction corresponds to a combination of muon and tau neutrinos,
which we denote $\nu_x$.  We also have $\omega_0=(m_1^2+m_2^2)/2E$,
the vacuum oscillation frequency is $\omega_{\bf p}=(m_2^2-m_1^2)/2E$
with the energy $E=|{\bf p}|$; ${\bf L}$ is a unit vector pointing in
the direction singled-out by the neutrino potential in the charged
fermion background, and $\lambda$ its normalization; for the cases
considered here, ${\bf L}={\bf e}_z$ and $\lambda=\sqrt2\,G_{\rm
  F}n_{e}$.  The unit vector ${\bf B}$ points in the mass eigenstate
direction in flavor space, such that ${\bf B}\cdot{\bf L}= -
\cos2\theta$, where $\theta$ is the vacuum mixing angle. The
neutrino-neutrino interaction couplings are given by $\{g_0,\,{\bf
  g}\}$, with the Standard Model case corresponding to $g_0=2$, $|{\bf
  g}|= 0$; it is thus the vector ${\bf g}$ which is characteristic of
NSI, its third component setting possible non-universal couplings
(\ie, $\nu_e-\nu_e$ different from $\nu_x-\nu_x$) while its first two
components characterize flavor-violating operators. Note that we have
assumed that the effective four-neutrino vertex is a good description
of the new dynamics. This is done having in mind an effective field
theory correction to the Standard Model dynamics induced by an energy
scale $M\gg M_Z$, with $|{\bf g}|\sim (M_Z/M)^2$.

With above definitions, the neutrino EOMs assume the form
\begin{equation}
 \dot{\bf P}_{\bf p} =\big(\omega_{\bf p}{\bf B}
 + \lambda {\bf L} +{\bf \Omega}^{S}_{\bf p}   \big) \times{\bf P}_{\bf p}\,,
\end{equation}
where the neutrino-neutrino interaction ``Hamiltonian'' ${\bf \Omega}^{S}_{\bf p}$ is 
\begin{widetext}
\begin{eqnarray}\label{eq:eom2}
 {\bf \Omega}^{S}_{\bf p}  = \mu\int {\rm d}{\bf q}
  \left(1-{\bf v}_{\bf q}\cdot{\bf v}_{\bf p}\right)\left\{
 \bigg[g_0 \xi +\bg\cdot ({\bf P}_{\bf q}-{\overline{\bf P}}_{\bf q})\bigg]\bg +\frac{1}{4}(g_0^2-|\bg|^2)
({\bf P}_{\bf q}-{\overline{\bf P}}_{\bf q})\right\} \,\ ,
\end{eqnarray}
\end{widetext}
and we have defined the parameter $\mu = \sqrt2\,G_{\rm F}n_{\bar\nu}$
which normalizes the neutrino-neutrino interaction strength.  For
antineutrinos, the EOMs are the same as for neutrinos with the
substitution $\omega_{\bf p}\to-\omega_{\bf p}$.  It is also trivial
to check that in the limit of Standard Model-only couplings ($g_0=2$, $|{\bf g}|=
0$), no direction is singled out in the flavor basis and one recovers
the standard neutrino-neutrino Hamiltonian.

In our analysis we will assume the single-angle approximation,
in which we substitute the angular structure of neutrino-neutrino interactions
in \eq~(\ref{eq:eom2}) 
 with an effective neutrino-neutrino interaction strength $\mu_r$ such that:
 \begin{equation}
\mu (1 - {\bf v}_{\bf p}\cdot {\bf v}_{\bf q})  \longrightarrow \mu_r  \, \ .
\end{equation}
This approximation has been numerically shown to be a good description
of the flavor evolution in the supernova environment, where a
significant asymmetry between neutrinos and antineutrinos is expected.
We have explicitly checked that the single-angle approximation is a
good description of the flavor evolution in presence of nonstandard
interactions also for the examples shown in this work.  In particular,
the asymmetry between neutrinos and antineutrinos prevents possible
effects of multi-angle decoherence in the neutrino
ensemble~\cite{EstebanPretel:2007ec}.  Inspired by this case, in our
numerical examples we use an effective neutrino interaction
strength~\cite{Hannestad:2006nj}
\begin{equation}  
 \mu_r = 0.35 \times 10^{5} \,\ \textrm{km}^{-1}\left[ 1-(1-r_{10}^{-2})^{1/2}\right]
 r_{10}^{-2} \,\ ,
\end{equation} 
where $r_{10}= r/(10~{\rm km})$. We consider a spherically symmetric
system in which neutrinos are emitted by a sphere at $R=10$~km. In
this situation, the only spatial variable that characterizes the
flavor evolution is the radial coordinate $r$.  Concerning the other
input parameters, in our numerical examples we will consider a single
mode system with $\omega= 0.3~{\rm km}^{-1}$, corresponding to
typical supernova neutrino energies for the atmospheric mass squared
difference. We choose $\theta=0$ which, in the standard case, would
mean that no flavor evolution can occur. We always assume $g_0=2$,
which essentially means that the overall strength of the
neutrino-neutrino interactions are given by the Standard Model. 

There is an important property which is worth noting: If $\bg
\parallel \bL$, then the quantity $\bB\cdot \bD = \bB \cdot ({\bf P}
-\bar{\bf{P}})$ is conserved.  This property is nothing but the
conservation of lepton number found in the standard
case~\cite{Hannestad:2006nj,Duan:2007mv}.  In this situation, one
expects to recover the standard dynamics. In the following, we shall
thus concentrate on the cases where $\bg$ is not parallel to $\bL$,
\ie, when flavor-changing currents are present. In the standard case,
where $\bB\cdot \bD = -D_z$ (for $\theta=0$) is preserved, to get a
diagnostic of the system it is enough to follow \eg~the evolution of
$\bar{P}_z$, which describes the polarization state of
antineutrinos. In the present case, it is useful to monitor also
$D_z$, where the differences with respect to the Standard Model case manifest more
clearly.

\begin{figure*}[t]
\begin{center}
\includegraphics[width=0.75\textwidth,clip=true]{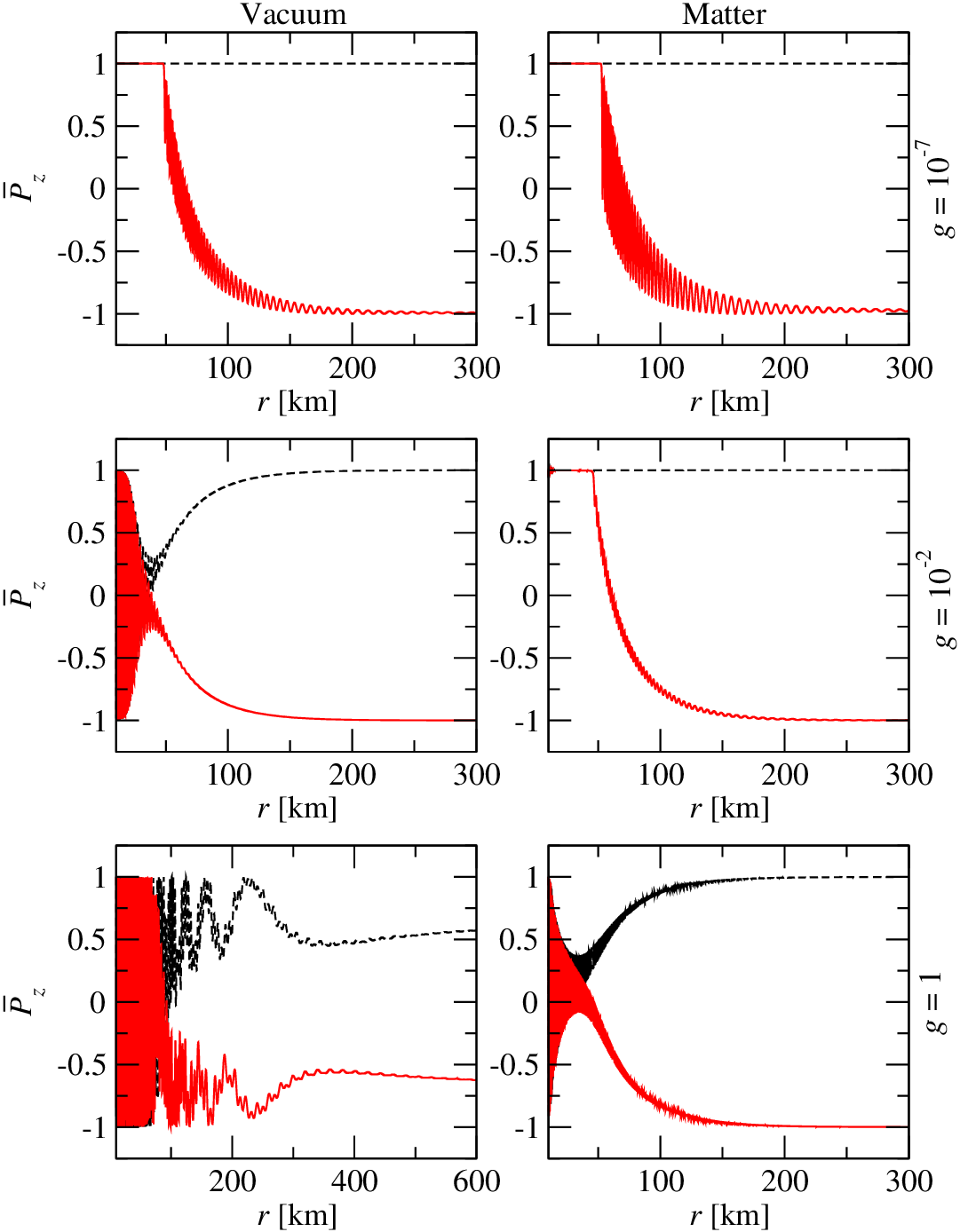}
\end{center}
\caption{The single-angle evolution of $\bar{P}_z$ in normal (dashed curves) 
and inverted hierarchy (solid curves) for $\theta=0$ in vacuum (left panels) and
in presence of matter (right panels). We consider values of nonstandard coupling
$g_1$ equal to $10^{-7}$ (top panels), $10^{-2}$ (central panels) and $1$ (bottom
panels). Note the different scale for $r$ in the lower left panel.}  \label{fig:pzbar}
\end{figure*}

\section{Flavor evolution }\label{SinAn}

\begin{figure*}[t]
\begin{center}
\includegraphics[width=0.8\textwidth,clip=true]{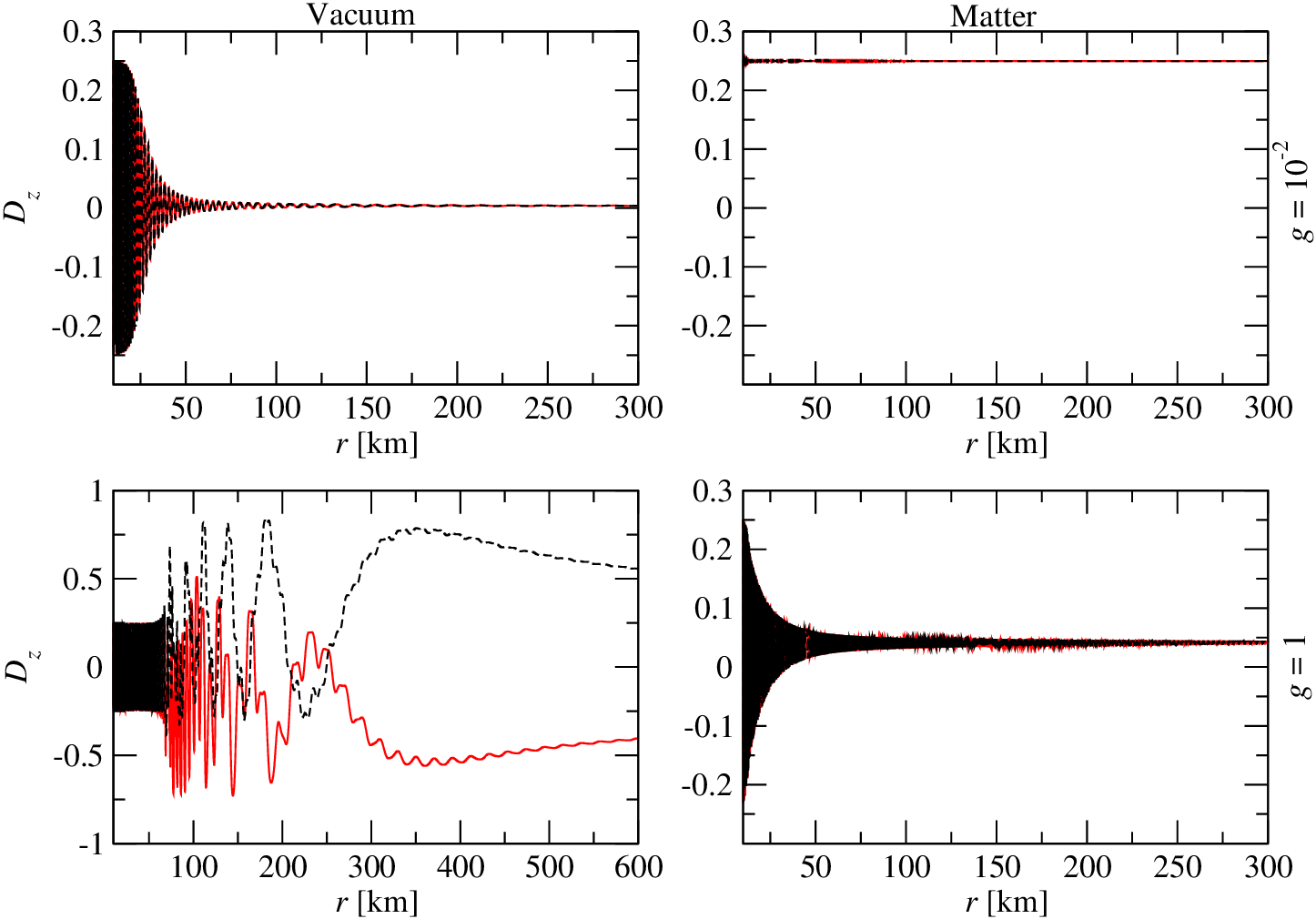}
\end{center}
\caption{The evolution of $D_z$ in normal (dashed curve) and inverted
(solid curve) in vacuum (left panels) and
in presence of matter (right panels). We use
$g_1 = 10^{-2}$ (upper panels) and $g_1 = 1$ (bottom
panels). Note the different scale for $r$ in the lower left panel.} \label{fig:Dz}
\end{figure*}


\subsection{Vacuum case}

To illustrate the effect of nonstandard neutrino-neutrino couplings on
the flavor evolution of a neutrino gas we first consider a case in
which the matter term $\lambda=0$. Since we have chosen $\theta=0$, we
assume that only $g_1 \neq 0$ without loss of generality.  
We start our discussion of NSI effects  referring to values of $|{\bf g}|\leq 10^{-1}$.  
In this regime, the terms ${\mathcal O}(|{\bf g}|^2)$ in \eq~(\ref{eq:eom2}) 
are sub-leading and it is clear that
the new terms have the form ${\bf g} \times {\bf P}_{\bf p}$.

The nonstandard couplings are expected to play a significant role when
the transverse nonstandard term $\xi \mu_r g_1$ in \eq~(\ref{eq:eom2})
dominates over the vacuum term $\omega B_z=-\omega$. Thus, we expect
NSI effects to be relevant for interaction strengths $g_1\gtrsim
10^{-4}$.  For smaller values, we expect to recover the standard
picture.  However, we note that also smaller values of nonstandard
coupling can have an interesting consequence. If we assume vanishing
vacuum mixing, we do not expect any flavor conversion in the standard
case, since the polarization vectors are exactly aligned with ${\bf
  B}$. However, in this situation, the nonstandard term transverse to
${\bf B}$ is enough to trigger an instability in flavor space,
producing a small offset with respect to the ${\bf B}$ direction. This
is enough to obtain flavor transformations in inverted hierarchy
analogous to the standard case~\cite{Hannestad:2006nj}. In
particular, we initially observe the synchronized oscillations in
which the polarization vectors essentially remain fixed to their
initial value until the start of the bipolar oscillations, which lead
to the inversion of the polarization vectors, conserving lepton flavor
number at each step.  In normal hierarchy the system is initially in
its stable configuration and the nonstandard effects have no impact.
The behaviour of ${\bar P}_z$ is represented in the top left panel of
\fig~\ref{fig:pzbar} for the case of $g_1=10^{-7}$.

For $g_1 \gtrsim 10^{-4}$ we observe that a transverse nonstandard
component is enough to produce wild oscillations in the system.
Essentially, the nonstandard interactions act as an external force on
the standard system, violating the symmetry that keeps $D_z$ constant.
Naively, it seems that the effect can be easily accounted for by going
into a frame co-rotating along
$\bg$~\cite{EstebanPretel:2008ni,Duan:2007mv}: we have just introduced
a nonstandard matter potential profile. The angle between $\bB$---now
rotating---and $\bg$ is not necessarily small if $\bg$ has a component
perpendicular to $\bB$, even for small vacuum mixing angle.  More
important, however, is that in this case the initial condition of the
neutrino state might be significantly offset compared with the ${\bf
  g}$ vector, \ie, ${\bf P}(0)$ and $\bar{\bf P}(0)$ might have a
significant component orthogonal to $\bg$, resulting in a large
amplitude of oscillation.

In presence of transverse nonstandard terms, the positive electron lepton number
we started with is not conserved, since
\begin{equation}
\dot D_z \simeq \mu_r \xi \left[g_1 D_y-g_2 D_x \right] \,\ .
\end{equation}
In \fig~\ref{fig:Dz} (top left panel) we show the behaviour of $D_z$
for the case of $g_1=10^{-2}$.  At small radii $D_z$ oscillates with a
full amplitude.  However, reducing $\mu_r$ the amplitude of the
oscillations diminishes: The decline of the upper envelope of $D_z$
follows the decrease of $g_1 \mu_r$ compared to the vacuum oscillation
frequency $\omega$.  Referring to the ``pendulum''
analogy~\cite{Hannestad:2006nj,Duan:2007mv}, the external force
introduced by the NSI is initially strong enough to completely
dominate the dynamics. However, as the strength $g \mu_r$ of the
external force decreases, it is finally so much weaker than the
potential $\omega$ that it can no longer lead to significant
oscillations. If the NSI term is sufficiently strong, the excess in
electron neutrinos we started with will relax to complete flavor
equilibration.

\begin{figure}[t]
\begin{center}
\includegraphics[width=0.9\columnwidth,clip=true]{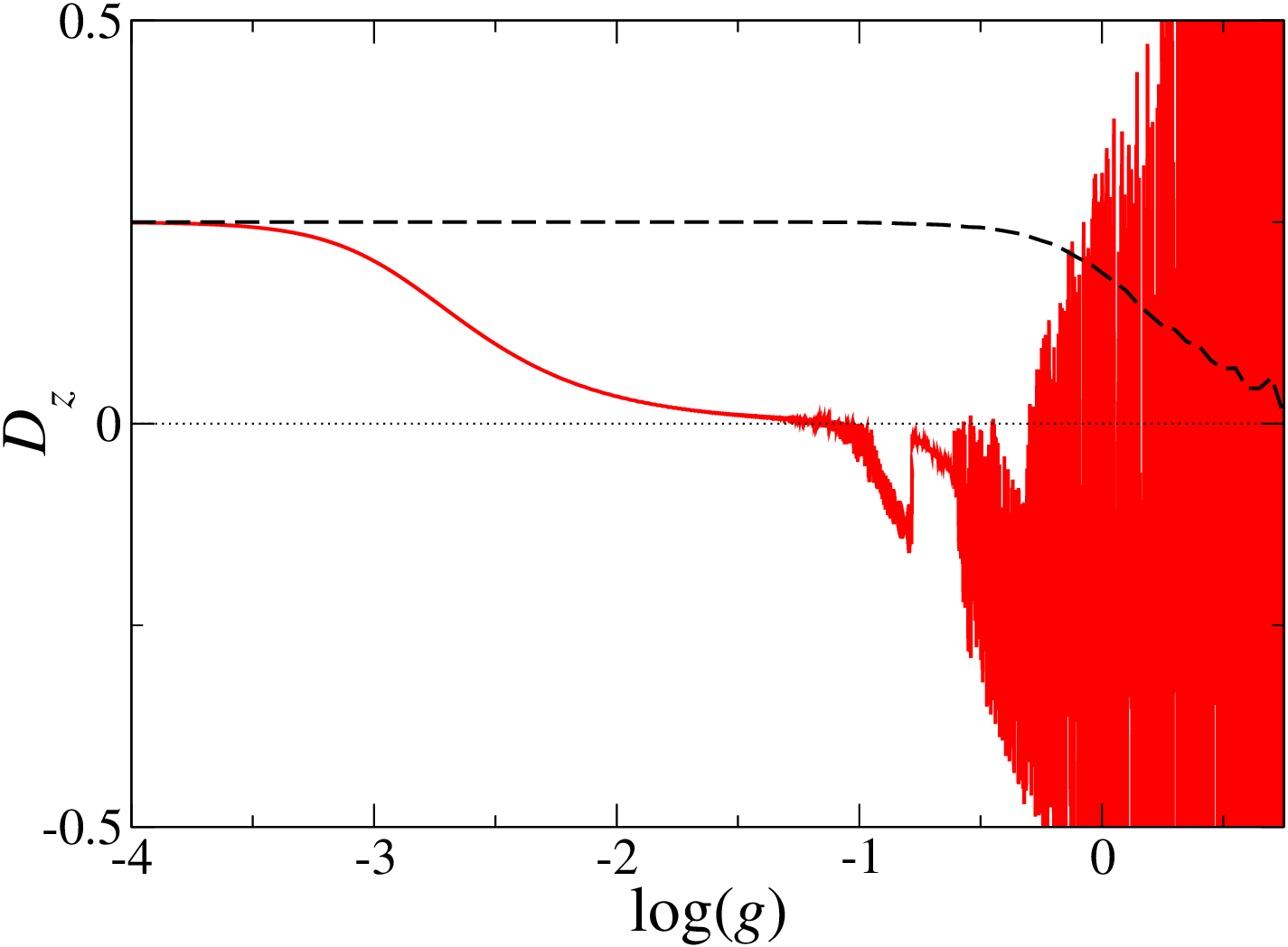}
\end{center}
\caption{The value of $D_z$ at the end of flavor evolution as a
  function of $g_1$ for inverted neutrino mass hierarchy in the vacuum
  case (solid curve) and in presence of matter (dashed curve). 
 Similar to the normal mass hierarchy case (not shown).} \label{fig:Dzvsg}
\end{figure}

 In \fig~\ref{fig:Dzvsg}, we show the behaviour of $D_{z}$ at the
 end of the flavor conversions as a function of $g_1$. We observe that
 nonstandard effects start to arise for $g_1 \gtrsim 10^{-4}$, in
 agreement with our prediction. For $10^{-2}\lesssim g_1 \lesssim
 10^{-1}$ we see that $D_z$ saturates at zero. The corresponding
 behaviour of $\bar{P}_z$ is shown in \fig~1 for $g_1=10^{-2}$
 (central left panel) for both normal and inverted hierarchy.  At the
 beginning, it oscillates with full amplitude which is then reduced 
 following the decrease of $\mu_r$. When the NSI effects have
 saturated and $D_z$ has reached zero, the polarization vectors
 follow the standard ``pendulum'' dynamics with null lepton
 number~\cite{Hannestad:2006nj}. As $\mu_r \to 0$, the kinetic energy
 of the flavor pendulum tends to zero and in both normal and
 inverted hierarchy the polarization vectors tend to minimize the
 potential energy~\cite{Hannestad:2006nj}
\begin{equation}
E_{\omega} = \frac{\omega}{2}\left(|{\bf S}| + {\bf B}\cdot {\bf S} \right) \,\ ,
\end{equation}
where ${\bf S}={\bf P}+\bar{\bf P}$. In normal neutrino mass
hierarchy, where $\omega {\bf B}$ points in the negative
$z$-direction, $\bar{P}_z$ rises again and returns to its initial
value.  In inverted hierarchy, where ${\bf B}$ points in the positive
$z$-direction and the original equilibrium position was unstable, the
initial dynamics are enough to trigger significant flavor changes. The
resulting behaviour is the total inversion of ${\bar P_z}$. The main
effects of a sub-leading nonstandard neutrino-neutrino interaction is
thus to produce a flavor equilibration in the ensemble.

We now consider large values of the ${\bf g}$ couplings. Even if from
a theoretical perspective these are quite unnatural, the loose
experimental bounds on these interactions allow us to play also with
these wilder scenarios.  Large couplings produce a modulation of the
standard neutrino-neutrino term in \eq~(\ref{eq:eom2}). Note that in
the extreme case of $|{\bf g}|=2$, the NSI term exactly compensates
the standard interaction term. If it were not for the other
nonstandard term on the right hand side of \eq~(\ref{eq:eom2}), the
system would become linear.  This term produces a precession of the
polarization vector ${\bf P}$ around the fixed ${\bf g}$ direction in
flavor space, with a speed modulated by $\mu_r {\bf g} \cdot {\bf D}$.
For generic large values of the NSI couplings, the motion of the
polarization vectors in flavor space is thus a combination of two
precessions, one around ${\bf D}$ and the other around ${\bf g}$, with
comparable speed.  Under this condition, the final value of $D_z$ at
the end of the evolution is quite unpredictable as shown in
\fig~\ref{fig:Dzvsg}. It is determined by the detailed combination of
the two different forces that drive the system: in particular, for
$g_1 =$ few $10^{-1}$, it is mostly the linear non-standard term that
matters, while for larger couplings the quadratic corrections are
important.  The behavior of $\bar{P}_z$ and $D_z$ as functions of $r$
is shown in \figs~\ref{fig:pzbar} and \ref{fig:Dz} (lower left
panels), respectively, for the case of $g_1=1$ in both normal and
inverted hierarchy.  Initially, we observe wild oscillations in these
vectors that decrease their frequency following the decline of
$\mu_r$. The evolution seems to maintain no track of the standard
behaviour. The nice pendulum analogy, driving the polarization
vectors in different directions for the different neutrino mass
hierarchies, remains. However, the fast oscillations of the final
value of $D_z$ as a function of $g_1$ inhibits the ability of making
precise predictions.

\subsection{Effect of Matter Background}

\begin{figure}[t]
\begin{center}
\includegraphics[width=0.9\columnwidth,clip=true]{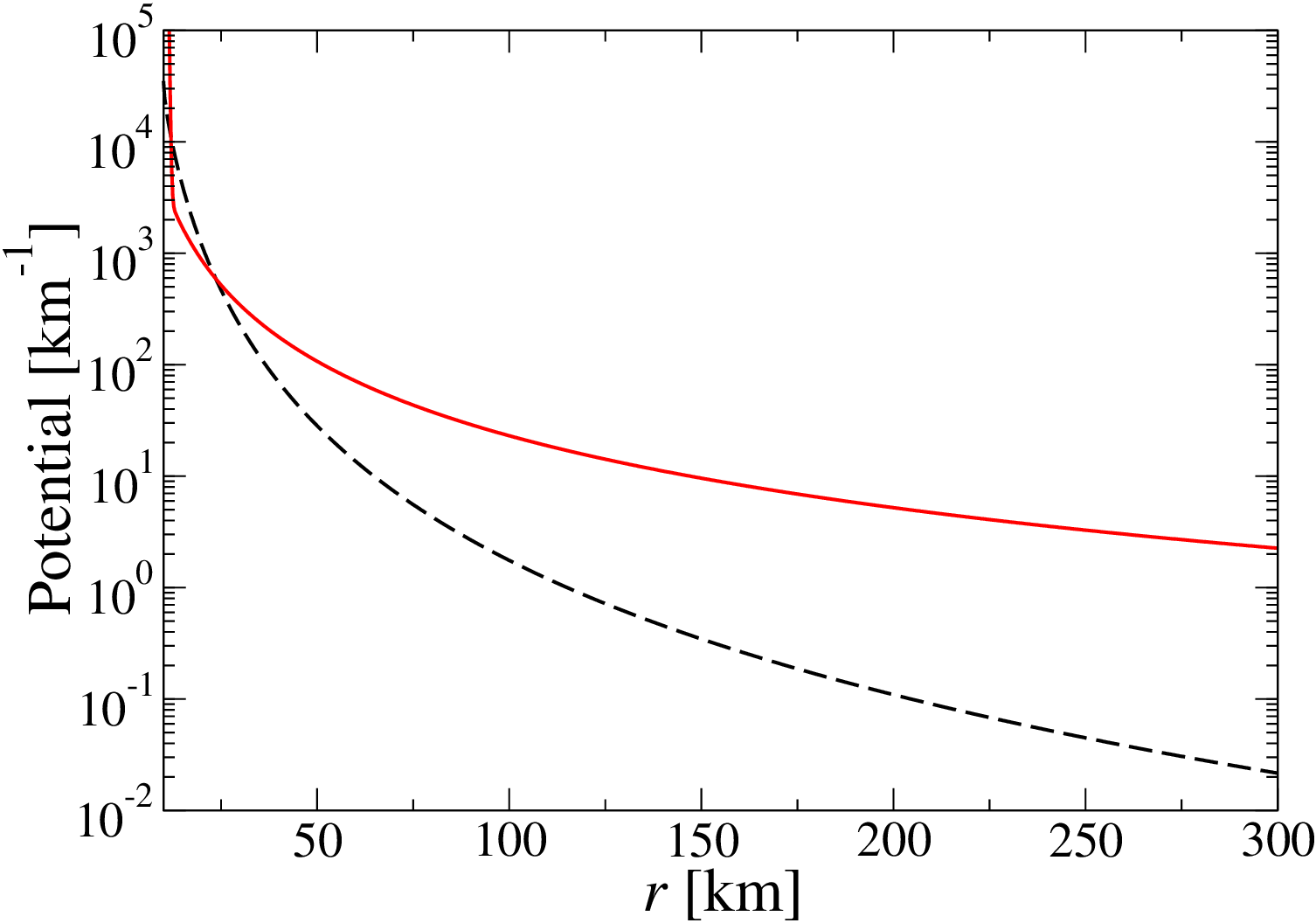}
\end{center}
\caption{Radial profile for the matter potential $\lambda$ (solid
  curve) and for the neutrino-neutrino potential $\mu_r$ (dashed
  curve).} \label{fig:potential}
\end{figure}
 
In a realistic supernova environment, the propagation of a dense
neutrino gas is also affected by the ordinary matter background.  To
characterize the matter potential $\lambda$ we refer to the
time-dependent parametrization used in
\Ref~\cite{Fogli:2003dw,Fogli:2007bk}, inspired by supernova
shock-wave simulations. In the following, we will focus on late times,
fixing $t=7$~s after the core bounce. In \fig~\ref{fig:potential}, we
show the $\mu_r$ and $\lambda$ potentials for our input choices.
 
For typical supernova conditions, the ordinary matter term $\lambda
{\bf L}$ dominates over sub-leading nonstandard interaction terms.  The
effect of a dominant dense matter term is to project the EOMs along
the weak-interaction direction~\cite{EstebanPretel:2008ni}. In the
co-rotating frame around ${\bf L}$, the fast-rotating ${\bf g}$
transverse component is enough to trigger the flavor conversions in
inverted hierarchy in the case of ${\theta}=0$, but otherwise plays no
crucial role.

In particular, for our input choices, in \fig~\ref{fig:Dzvsg} we
observe that the matter effect guarantees the conservation of $D_z$
for $g_1 \lesssim 0.3$.  For smaller values of $g_1$, the behavior of
${\bar P}_z$ and $D_z$ as functions of $r$ is analogous to the
standard case as shown in \figs~\ref{fig:pzbar} and \ref{fig:Dz}
(upper right panels), respectively, for the case of $g_1=10^{-2}$.  If
we allow for $g_1 \sim {\mathcal O}(1)$, it can dominate over the
$\lambda$ term close to the neutrinosphere.  The behaviour of ${\bar
  P}_z$ and $D_z$ as functions of $r$ for $g_1=1$ is shown in
\figs~\ref{fig:pzbar} and \ref{fig:Dz} (lower right panels),
respectively.  We observe that, for $g_1=1$, the vectors initially
undergo very rapid oscillations. However, when $\mu_r g_1 < \lambda$,
the standard matter potential dominates the dynamics and the
oscillations only occur around the flavor direction, thus decreasing
the flavor oscillation amplitude. Thus, at the end of flavor
evolution, the value of $D_z$ saturates.

\section{Conclusions and Outlook}\label{concl}

In this work we have investigated the impact of nonstandard
neutrino-neutrino interactions on the flavor evolution of a dense
neutrino gas.  First, we have generalized the equation of motion for a
neutrino ensemble in presence of nonstandard neutrino-neutrino
interaction terms.  Then, we have shown that if ordinary matter
effects are sub-dominant, flavor-changing neutrino-neutrino couplings
lead to a dynamics very similar to the Standard Model analogue, with
the additional feature of a complete flavor equilibration of the
neutrino ensemble, both in normal and in inverted hierarchy. Including
matter effects of the magnitude expected in a realistic supernova
environment, sub-leading NSI corrections are suppressed by the
dominant dense matter effect and a picture similar to the
``standard'' one emerges. Small NSI thus have negligible effects, except
for triggering flavor conversions in inverted hierarchy even in the
limiting case of $\theta=0$.  All in all, we conclude that the
phenomenon of collective neutrino flavor evolution in supernovae is generally
robust with respect to sub-leading neutrino-neutrino interactions of the
four-fermion type. Potentially large effects can arise for values of the
nonstandard neutrino-neutrino couplings comparable to the ordinary
ones, which are phenomenologically allowed but  theoretically disfavored.

Still, to make the problem tractable we made several approximations
compared to a realistic physical system. Let us conclude this section by speculating 
on possible consequences of our findings in more realistic environments.
For example, in some situations the
ordinary matter effect might be weak enough for sub-leading NSI
effects to affect the dynamics. Perhaps the most notable example is in
O-Ne-Mg SN progenitors with rapidly decreasing density profiles, where
in particular the neutronization burst might be
affected~\cite{Duan:2007sh,Dasgupta:2008cd,Lunardini:2007vn}. Also, 
peculiar time-dependent signatures in the supernova neutrino burst could be
produced by non-negligible NSI effects, since, at least at sufficiently late time after 
core bounce, the
nonstandard neutrino-neutrino term could dominate over the dense
matter term in the low-radii region. Another physically interesting environment
is provided by accretion disks formed by
coalescing neutron stars,  where emitted neutrinos propagate in a region of relatively low
matter density along the disk axis~\cite{Dasgupta:2008cu}.
Nonstandard interactions could play some role in this context, perhaps
with an impact on the gamma-ray burst production by neutrino-neutrino
annihilations in such engines.  Also note that a  more realistic treatment 
of the problem  might require a detailed three-flavor study, which goes beyond our
present purpose.  

Further directions of investigation may be related to the presence of
nonstandard neutrino-charged fermion interactions. Their impact on
matter transitions in supernovae has been explored in different
works. In particular, in \Ref~\cite{EstebanPretel:2007yu} it has been shown that these could induce
new effects just above the neutrino sphere. Their impact on the
collective neutrino flavor evolution still remains to be
characterized.

Finally, we would like to remark that this study assumes that the
nonstandard effects do not affect the formation of the neutrino
spectra and the dynamics in the supernova core. It has been argued
in~\cite{Amanik:2004vm,Amanik:2006ad} that sufficiently large NSI of
neutrinos with SN matter may lead to drastic changes in the core
electron fraction and neutrino spectra. Since the NSI considered here
do not enter directly into the weak equilibrium, nor enhance
neutrino-nucleus collisions, one would naively guess that at least for
sufficiently small $g$ their effects should be sub-leading.  Of
course, a definitive proof of this working hypothesis would require
detailed supernova simulations, including nonstandard interaction
effects in the dynamics of the core-collapse.


\begin{acknowledgments}
We would like to thank Georg Raffelt for reading the manuscript and
giving useful suggestions and Walter Winter for comments. 
A.M.~also wishes to thank Basudeb Dasgupta for interesting discussions on this topic.
This work was supported in part by the Swedish Research Council
(Vetenskapsr\aa{}det) through Contract No.~623-2007-8066 [M.B.] and
Istituto Nazionale di Fisica Nucleare (INFN, Italy) [A.M.]. 
\end{acknowledgments}

\end{document}